\documentclass[twocolumn]{aastex631}
\usepackage[utf8]{inputenc}
\usepackage{CJK}  
\usepackage{graphicx} 
\usepackage{txfonts}
\usepackage{natbib}
\usepackage{hyperref}
\hypersetup{colorlinks}
\usepackage{multirow}
\usepackage{longtable}

\revised{\today}

\graphicspath{{fig9/}} %
\shorttitle{The BT algorithm at work: A2029}
\shortauthors{Yu \& Diaferio}

\begin{document}

\title{The Blooming Tree Algorithm at Work: Clusters, Filaments and Superclusters in the Field of A2029}
\correspondingauthor{Heng Yu}
\email{yuheng@bnu.edu.cn}

\author[0000-0001-8051-1465]{Heng Yu}\
\affil{School of Physics and Astronomy, Beijing Normal University, Beijing, 100875, China}

\author[0000-0002-4986-063X]{Antonaldo Diaferio}
\affiliation{Dipartimento di Fisica, Universit\`a di Torino, Via
  P. Giuria 1, I-10125 Torino, Italy}
\affiliation{Istituto Nazionale di Fisica Nucleare (INFN), Sezione di Torino, Via
  P. Giuria 1, I-10125 Torino, Italy}

\begin{abstract}
The Blooming Tree (BT) algorithm, based on the hierarchical clustering method, is designed to identify clusters, groups, and substructures from galaxy redshift surveys. We apply the BT algorithm to a wide-field ($10\times 10$ deg$^2$) spectroscopic dataset centered on the galaxy cluster A2029. The BT algorithm effectively identifies all the X-ray luminous clusters and most of the optical clusters known in the literature, numerous groups, and the filaments surrounding the clusters, associating a list of galaxy members to each structure.
By lowering the detection threshold, the BT algorithm also identifies the three superclusters in the field. The BT algorithm arranges the clusters and groups that make up the superclusters in a hierarchical tree according to their pairwise binding energy: the algorithm thus unveils the possible accretion history of each supercluster and their future evolution.  
These results show how the BT algorithm can represent a crucial tool to investigate the formation and evolution of cosmic structures on non-linear and mildly non-linear scales. 
\end{abstract}


\section{Introduction} \label{sec:intro}

Hierarchical clustering is a popular unsupervised learning algorithm for identifying groups of objects with common properties.
It arranges all the objects in a dataset into a binary tree according to a specified similarity measure or metric. The graphic representation of the tree is called a dendrogram. The dendrogram has the distinct advantage of visualizing the hierarchical inner structures of the dataset. With a given tree-trimming threshold, the tree can be separated into groups \citep[see, e.g.,][for a detailed description]{everitt2011,2022Yu}.

Hierarchical clustering was first introduced into astronomy by \citet{Materne1978} for the identification of nearby groups and clusters of galaxies.
This pioneering effort established hierarchical clustering as the first astronomical three-dimensional group analysis method and sparked the interest of other researchers \citep{Tully1980,Tully1987,Gourgoulhon1992}.

 Although the Friends-of-Friends (FOF) algorithm became popular due to its simplicity
\citep{1976Turner,1982Huchra,2000Mahdavi,2007Yang,2009Finoguenov,2011Robotham,2016Tempel},
efforts have been made to explore the potential of hierarchical clustering for the identification of structures.
\citet{Serna1996} found that the projected binding energy, derived from the galaxy celestial coordinates and the line-of-sight velocity, is a sensible metric for cluster identification.
\citet{1999Diaferio} and \citet{Serra2011} proposed a physically motivated procedure to trim the dendrogram, based on the plateau of the galaxy velocity dispersion $\sigma$  along the main branch of the hierarchical tree.
This approach has been proven to be an effective procedure for extracting the members of a galaxy cluster \citep{2013Serra} and its substructures \citep{Yu2015, 2016Yu, Liu2018b}.

Based on these features, \citet{Yu2018a}  developed a new approach, the Blooming Tree (BT) algorithm, to identify structures more extensively. Rather than identifying the main branch of the tree, which is appropriate for the identification of the dominant structure in the field, the BT algorithm explores all the branches of the tree to look for key nodes that, by satisfying a specific criterion, identify all the galaxy systems in the field.
With this approach, we can identify all the sufficiently dense structures in the field without assuming spherical symmetry or hydrostatic equilibrium.

The BT algorithm belongs to the family of hierarchical clustering techniques. These techniques are not yet grounded in solid mathematical theorems. Therefore, the structures and
memberships identified by these technique depend on the arbitrary choice of their parameters. Consequently, the adoption of any particular technique is generally based on its performance on mock catalogs, where the actual properties of the systems can be compared with the properties of the systems identified by the technique. The BT algorithm is a substantial refinement of the $\sigma$-plateau and the caustic methods \citep{1999Diaferio, Serra2011, 2013Serra}. All these techniques were extensively  tested on mock galaxy redshift catalogs. When applied to galaxy clusters, the caustic technique yields a completeness, the fraction of identified true members, $f_c \sim 95\%$, within $3r_{200}$, where $r_{200}$ is the radius within which the cluster density is 200 times the critical density of the universe; the contamination, the fraction of interlopers in the observed catalog of members, drops from $f_i\sim 8\%$ at $3r_{200}$ to $f_i\sim 2\%$ at $r_{200}$ \citep{2013Serra}. No other technique for the identification of the members of a galaxy cluster provides such large completeness and small contamination at these large radii.  Similarly, \citet{Yu2018a} show that, when applied to  mock redshift surveys with 200 galaxies within 6 $h^{ -1}$ Mpc from the cluster center, the BT algorithm recovers $\sim 80\%$ of the real substructures and $\sim 60\%$ of the surrounding groups; in $\sim 60\%$ of the identified structures, at least $60\%$ of the member galaxies of the substructures and groups belong to the same real structure. Therefore, the BT algorithm currently is among the most reliable technique to identify galaxy systems from spectroscopic redshift surveys.

In addition to structure identification, the hierarchy among these structures contained in the binary tree provided by the BT algorithm is a relevant piece of information to investigate the dynamics of the galaxy systems.
This property of the algorithm is an invaluable advantage for the analysis of systems that are out of dynamical equilibrium such as superclusters.
Superclusters are generally defined as large gravitationally linked systems, composed of two or more galaxy clusters; they are identified as local overdensities in the galaxy distribution
\citep{1983OOrt, 1988Bahcall}.
Because our universe appears too young for superclusters to be virialized, they do not have regular shapes or clear boundaries.

The current supercluster identification is carried out mainly with the FoF algorithm applied to the Abell cluster catalog \citep{1976Rood, 1984Bahcall, 1985Batuski, 1989West, 1993Zucca, 1995Kalinkov, 2001Einasto}, the 2-degree Field \citep[2dF,][]{2007Einasto}, the Sloan Digital Sky Survey \citep[SDSS,][]{2011Luparello, 2012Liivam}, the X-ray ROSAT All-Sky Survey \citep[RASS,][]{2021Bohringer}, the eROSITA Final Equatorial-Depth Survey \citep[eFEDS,][]{2022Liu}, and the SRG/eROSITA All-Sky Survey \citep[eRASS1,][]{2024Liu}, among others.

Here, we apply the BT algorithm to
a catalog of galaxy spectroscopic redshifts in a wide field centered on the cluster A2029
to detect and analyse galaxy clusters and superclusters.
A2029 is a massive cluster ($M_{200}$ = 8.5 $\times$ 10$^{14}$ $M_\odot$) at redshift  $z=0.0773$ \citep{2019Sohn}. It is also the central cluster of a supercluster \citep{2001Einasto, 2012Liivam, 2014Chow}. There are a number of surrounding clusters and groups falling onto A2029 \citep{2012Walker, 2019Sohn, 2019Sohnb}.
Such a complex environment makes this field an ideal region for the application of the BT algorithm and for exploring the hierarchy of the galaxy systems.

We describe the BT algorithm in Section \ref{sec:method}; we present the data and the analysis in Section \ref{sec:data},  and the cross-check between our results and other cluster catalogs in Section \ref{sec:resu}. Section \ref{sec:superclusters} illustrates the properties of the superclusters and the large-scale filaments identified in the field, and Section \ref{sec:end} contains the summary and discussion.
Our analysis assumes a $\Lambda$CDM cosmological model with  parameters $H_0$ = 67.4 km~s$^{-1}$ Mpc$^{-1}$, $\Omega_m = 0.315$, and $\Omega_\Lambda = 0.685$ \citep{planck2018}.

\section{Method}
\label{sec:method}

The basic idea of the BT algorithm is to arrange all the objects in the data set in a binary tree 
according to the pairwise gravitational binding energies of these objects: pairs of objects with decreasing value of the binding energy will appear at deeper levels of the tree. By trimming the binary tree with an appropriate threshold, 
we can associate tree branches to well-defined kinematic structures.

The pairwise binding energy combines an estimate of the 
gravitational potential energy and of the kinetic energy derived from the three observable quantities, namely the two coordinates on the sky and  the light-of-sight velocity component:
\begin{equation}
E_{ij}=-G{m_i m_j\over R_p}+{1\over 2}{m_i m_j\over m_i+m_j}\Pi^2 \;,
\label{eq:pairwise-energy}
\end{equation}
where $R_p$ is the pair projected separation, which depends on the
angular diameter distances of the two galaxies, 
and $\Pi$ is the proper
line-of-sight velocity difference (see \citealt{Serra2011} for further details). 
Equation (\ref{eq:pairwise-energy}) clearly is only a proxy of the actual pairwise binding energy, because, out of the six phase-space coordinates of the galaxies, $E_{ij}$ includes only the three measurable coordinates. Any analysis based on eq. (\ref{eq:pairwise-energy}) can thus be affected by inaccurate estimates of the pairwise binding energy. Nevertheless, when applied to mock catalogs extracted from N-body simulations, the BT algorithm recovers, on average, 60\% of the real groups surrounding galaxy clusters, and at least 60\% of the actual galaxy members of these groups \citep{Yu2018a}. Other work shows that this limited phase-space information can still provide meaningful information on the galaxy-environment connection, including the galaxy infall time \citep{Oman2013} and the tidal mass loss of the falling galaxy \citep{Rhee2017}. 

For the galaxy mass, we adopt the typical mass of a luminous galaxy
$m_i = m_j = 10^{12}h^{-1}$ M$_\odot$ \citep{1999Diaferio}.
Assuming the same mass for all the galaxies irrespectively of their luminosity appears a simplistic assumption. However, the attempt of assigning a mass proportional to the galaxy luminosity implies a number of assumptions and parametrizations that would complicate the separation of the role of the adopted mass-luminosity relation from the role of the galaxy kinematics in the determination of the performance of the BT algorithm. In a possible procedure to adopt a mass spectrum, we would first infer the stellar mass from the galaxy luminosity based on an adopted mass-to-light ratio. This ratio depends on the galaxy morphology and luminosity, which, on turn, depends on the  aperture of the photometric measurement; even assuming that the morphological and photometric information is uniform over the entire spectroscopic sample, this aperture is generally fixed in angular size, so the luminosity of different galaxies correspond to the luminosity within different physical radii, and an additional correction procedure would be necessary. Finally, the BT algorithm requires the galaxy dynamical mass, so one needs a relation between the stellar mass estimated from the mass-to-light ratio and the mass of the galaxy dark matter halo.  This relation, estimated in N-body simulations, shows a substantial spread and it is correlated with the galaxy environment \citep[e.g.,][]{Mercado2025}. This outlined procedure shows the questionable complications that would be introduced by assigning a mass to individual galaxies based on their luminosity. We prefer to keep the procedure simple by assigning the same mass to all the galaxies. The effectiveness of this choice is confirmed by the satisfactory performance of the BT algorithm when tested on N-body simulations \citep{Yu2018a}.

Because the typical velocity dispersion of the members of a galaxy cluster is of the order of 1000 km~s$^{-1}$,
corresponding to a redshift $z \sim$ 0.003, 
the peculiar velocities of galaxies in a cluster at redshift $\sim 0.02$  cause, on average, $\sim 15$\% uncertainty on the galaxy distance.
So we limit our sample to galaxies at redshift $z>0.02$.

We perform the hierarchical clustering analysis by building a binary tree as follows 
(see \citealt{1999Diaferio,Serra2011,Yu2015} for further details): 

\begin{itemize}
\item[i.] initially each galaxy is in a group $G_\alpha$, of which it is the only member;
\item[ii.] we estimate the binding energy between two groups $G_\alpha$ and $G_\beta$ as $E_{\alpha\beta}={\rm min}\{E_{ij}\}$,
	where $E_{ij}$ is the pairwise binding energy between the galaxy $i\in G_\alpha$ and the galaxy $j\in G_\beta$; 
\item[iii.] the two groups with the smallest binding energy are
replaced with a single group and the total number of groups is decreased by one;
\item[iv.] we repeat the procedure from step (ii) until only one group is left.
\end{itemize}

When the tree is built, we perform the BT procedure to extract the structures.
The first step is to find the local minima of the binding energy of the leaves of the tree. We call these local minima "buds". We then identify the branches of the tree where the buds are located and calculate the properties of each node on the branches.
As tested and discussed in \citet{Yu2018a}, we combine these properties in the quantity
\begin{equation}
\eta_{sub} = \frac{\sigma_v}{d_{avg}~\sqrt{n} } {\rm km ~s^{-1} Mpc^{-1}},
\end{equation}
where $\sigma_v$ is the velocity dispersion of the leaves hanging from the node, $n$ is the number of leaves,
and $d_{avg}$ is the average distance between the leaves: 
As shown by \citet{Yu2018a} (their Eqs. 2-5), $\eta_{sub}$ is an indicator of the mass density of the candidate structure identified by the node.
To identify the final structures, in other words to identify the buds that actually bloom, we consider the quantity $\Delta\eta=\eta_{sub}-\eta_{bck}$, where $\eta_{bck}$ is estimated for  the background region
surrounding the structure; specifically, the surrounding region is centered on the candidate structure and
extends to a projected radius five times
larger than the radius of the structure, namely the
radius of the smallest circle enclosing the entire structure on the
plane of the sky. The adoption of $\Delta\eta$ compensates for the inhomogeneities of the galaxy distribution  caused by the completeness fluctuation of the surveys. The criterion $\Delta\eta$ is the only free parameter set by the user in most cases.
Generally speaking, the larger is the threshold $\Delta\eta$, the more compact are the identified structures. 

The Python source code of the whole procedure is available at the Astrophysics Source Code Library (ASCL)\footnote{\url{https://ascl.net/2503.032}}, Zenodo\footnote{\url{https://doi.org/10.5281/zenodo.15479906}}
and the GitHub\footnote{\url{https://github.com/henrysting/blmtree}}.

\section{Data}
\label{sec:data}

A2029 is a nearby cluster with substantial spectroscopic data available in the literature. 
We compile a catalog of 26406 spectroscopic redshifts in the field of 10 $\times$ 10 deg$^2$ centered on A2029 (RA$=228.0$ deg, DEC$=5.5$ deg),
corresponding to an area of 53 $\times$ 53 Mpc$^2$
at the redshift of A2029, $z=0.0773$. The field has a radius $\sim 14R_{200}$, with $R_{200}=1.91$ Mpc the radius of A2029  \citep{2019Sohn,2019Sohnb}.
There are 26158 spectroscopic galaxy-type redshifts from SDSS DR18 \citep{2023dr18},
185 redshifts from LAMOST DR9 \citep{2012RAA....12..723Z},  37 from  2MASS \citep{2012ApJS..199...26H}, and 26 from NASA/IPAC Extragalactic Database \citep[NED,][]{NED}.

\begin{figure}
 \centering
 \includegraphics[trim={0 3cm 0 3cm},width=0.47\textwidth,]{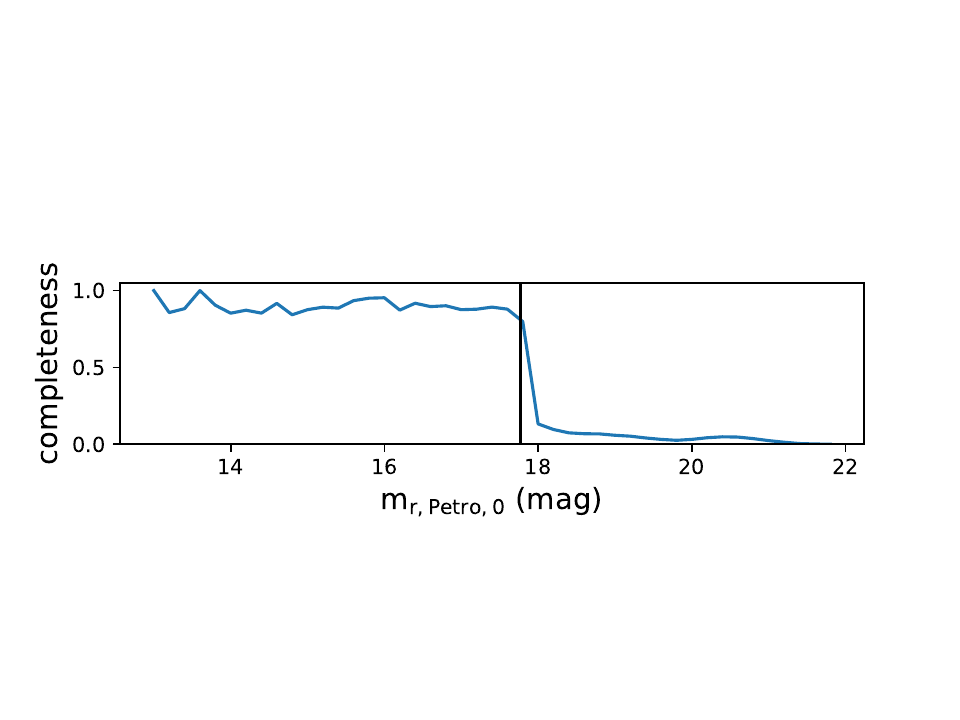}\\
 \includegraphics[width=0.49\textwidth]{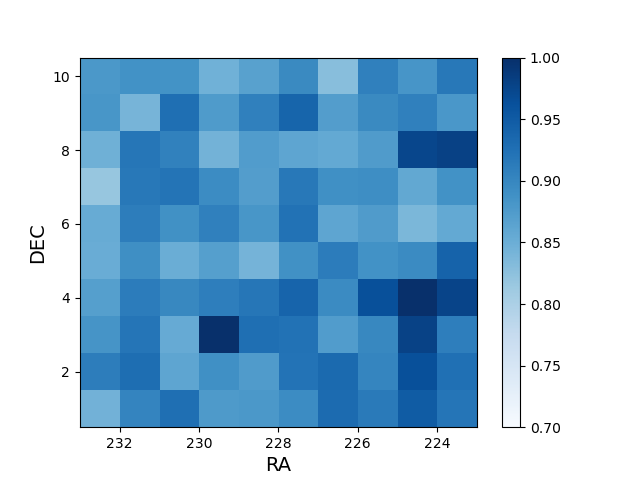}
 \caption{The top panel is the spectroscopic completeness 
as a function of the r-band magnitude. The solid vertical line indicates $m_{\rm r,Petro,0} = 17.77$.
The bottom panel is the two-dimensional completeness map of the A2029 field.}
 \label{fig:comp2d}
\end{figure}

With the photometric catalog of SDSS DR18, we estimate the spectroscopic completeness
as a function of magnitude $m_{\rm r,Petro,0}$.
For 12848 galaxies with $m_{\rm r,Petro,0} \le 17.77$
\citep{2002Strauss}, the overall completeness reaches 89.3\%. The two-dimensional cumulative completeness map is shown in Fig. \ref{fig:comp2d}. In this 10$\times$10 mosaic map, the completeness of each pixel is larger than  81\% with 4\% standard deviation.

\citet{2019Sohn} provide a much deeper spectroscopic data, with completeness larger than 90\% at $m_{\rm r,Petro,0}$ = 20.5 mag, on an area $\sim 40\times 40$ arcmin$^2$ centered on A2029. We do not include this data set in our sample to preserve the completeness uniformity of our catalog and to avoid the identification of an artificial overabundance of structures   in the central region.

\begin{figure}
 \centering
 \includegraphics[width=0.49\textwidth]{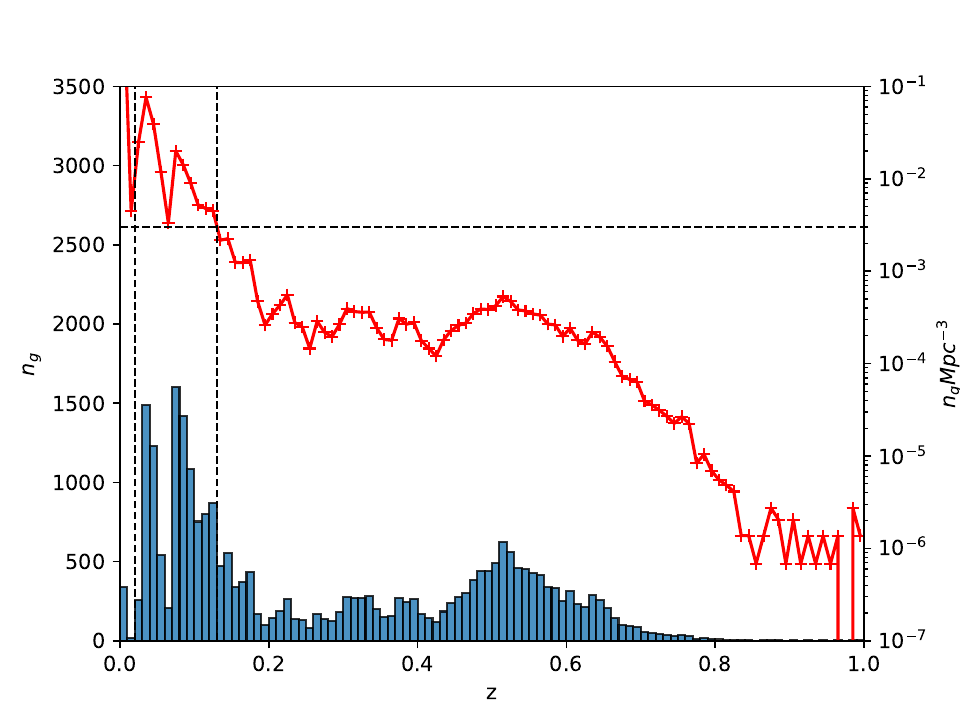}
 \caption{The redshift distribution of the galaxies in our sample (solid histogram). The galaxy number density in each redshift bin is represented by the red curve. The two dashed vertical lines indicate redshifts 0.02 and 0.13. The horizontal dashed line indicates the galaxy number density  0.003 Mpc$^{-3}$. }
 \label{fig:z1d}
\end{figure}

The redshift distribution of our catalog is shown in Fig. \ref{fig:z1d}.  We estimate the galaxy number density in each redshift bin as $n_{g}/[\Omega D_{A}^2(z) l_{bin}]$, where $n_g$ is the galaxy number in the redshift bin, $\Omega$ is the solid angle subtended by the 10 $\times$ 10 deg$^2$ sky area of the field,
$D_{A}(z)$ is the angular diameter distance, and $l_{bin}$ is the comoving width of the redshift bin. Figure \ref{fig:z1d} shows that the galaxy number density, indicated by the red curve, rapidly drops with increasing redshift.
Therefore, although there are known clusters located at high redshift in this field, for example, RX J1524.6+0957 at z=0.5,  MCXC J1524.0+1003 and MCXC J1524.8+1005 at z=0.2, the number of spectroscopic data at these redshifts is insufficient for a reliable application of the BT algorithm. We thus 
 limit our analysis to the redshift range 0.02-0.13, which represents a substantial proper volume around A2029.
There are 10258 galaxies in our catalog that fall within this redshift range. Our further analysis considers only this limited sample of galaxies.

\section{Structures}
\label{sec:resu}

The BT algorithm identifies structures according to the parameter $\Delta \eta$,
which quantifies the difference between the average mass density of the structure and its surrounding region: the larger is the threshold $\Delta\eta$, the more compact is the identified structures. To identify substructures in galaxy clusters, \citet{Yu2018a} adopt $\Delta \eta = 100$; they test this threshold on N-body simulations. Here, to identify clusters and groups of galaxies, we adopt $\Delta \eta = 20$. With this threshold, we identify structures  that are expected to be less dense than substructures within galaxy clusters.

\begin{figure*}
 \centering
 \includegraphics[width=0.96\textwidth]{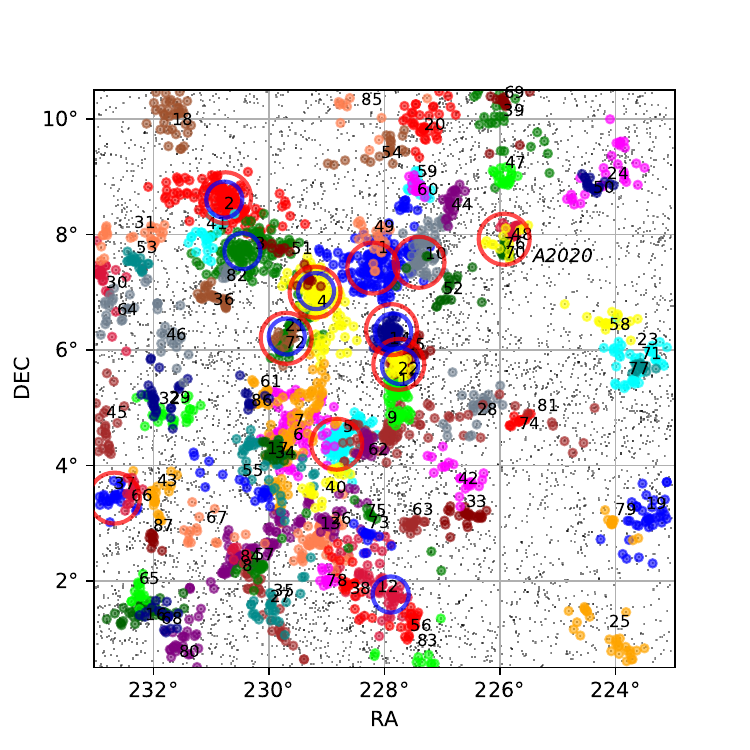}
 \caption{Distribution of the BT structures on the sky. Different structures are shown with different colors. However, some colors are reused. The identification numbers of the BT structures are placed near their centroids. The grey points, the red circles and the blue circles show the galaxies, the Abell clusters, and the MCXC clusters, respectively.}
 \label{fig:2dist}
\end{figure*}

The BT algorithm identifies 87 structures with velocity dispersion larger than 200 km~s$^{-1}$ and at least 10 members. These 
structures are listed in Table \ref{tab:subs} with the prefix ``g''. For each structure, the table lists the number of members $n_{\mathrm{mem}}$, the celestial coordinates of the geometric center, the redshift $z$, and the velocity dispersion $\sigma_v$.
Figure \ref{fig:2dist} shows the distribution of these structures on the sky.
We estimate the error $\Delta\sigma_v$ on the velocity dispersion $\sigma_v$  according to the relation
\begin{equation} 
{\rm log_{10}}(\Delta \sigma_v / \sigma_v) = -0.49 {\rm log_{10}}(n_{mem}) - 0.17 
\end{equation}
which is derived with a bootstrap method, as detailed in Appendix \ref{app:bootstrap}.

\begin{longtable}{|c|r|c|c|c|c|}
\caption{BT structures with threshold $\Delta \eta$ = 20}
\label{tab:subs}\\
\hline
 ID & $n_{\rm mem}$ & RA (deg) & DEC (deg) & $z$  & $\sigma_v$ \\
\hline
\endfirsthead

\multicolumn{6}{c}{\textbf{continued}} \\
\hline
 ID & $n_{\rm mem}$ & RA (deg) & DEC (deg) & $z$  & $\sigma_v$ \\
\hline
\endhead

\hline
\endfoot

g1 &  174 & 228.10541 & 7.47120 & 0.0452 & 491 $\pm$ 26 \\
g2 &  147 & 230.75081 & 8.60306 & 0.0347 & 712 $\pm$ 42 \\
g3 &  133 & 230.35705 & 7.70882 & 0.0450 & 535 $\pm$ 33 \\
g4 &  132 & 229.17687 & 6.98369 & 0.0347 & 573 $\pm$ 35 \\
g5 &  105 & 228.75443 & 4.39871 & 0.0974 & 595 $\pm$ 41 \\
g6 &  101 & 229.63550 & 4.52068 & 0.0367 & 406 $\pm$ 29 \\
g7 &   93 & 229.61360 & 4.54410 & 0.0487 & 595 $\pm$ 44 \\
g8 &   90 & 230.42393 & 2.44845 & 0.0854 & 776 $\pm$ 58 \\
g9 &   88 & 227.93879 & 4.51708 & 0.0364 & 401 $\pm$ 30 \\
g10 &   73 & 227.32172 & 7.63877 & 0.0768 & 430 $\pm$ 36 \\
g11 &   68 & 227.76212 & 5.25172 & 0.0805 & 477 $\pm$ 41 \\
g12 &   64 & 228.06140 & 1.87079 & 0.0387 & 306 $\pm$ 27 \\
g13 &   63 & 229.11558 & 2.77073 & 0.1117 & 480 $\pm$ 43 \\
g14 &   62 & 227.89391 & 6.26200 & 0.0799 & 571 $\pm$ 51 \\
g15 &   55 & 227.65662 & 5.81437 & 0.0786 & 560 $\pm$ 53 \\
g16 &   47 & 232.20079 & 1.46399 & 0.0749 & 392 $\pm$ 40 \\
g17 &   44 & 230.01420 & 4.09724 & 0.0523 & 278 $\pm$ 30 \\
g18 &   36 & 231.66258 & 10.09184 & 0.0759 & 356 $\pm$ 42 \\
g19 &   36 & 223.43918 & 3.07582 & 0.0278 & 286 $\pm$ 34 \\
g20 &   36 & 227.31262 & 9.90722 & 0.0828 & 314 $\pm$ 37 \\
g21 &   35 & 229.69962 & 6.18170 & 0.1031 & 452 $\pm$ 54 \\
g22 &   31 & 227.74490 & 5.73385 & 0.0735 & 262 $\pm$ 33 \\
g23 &   31 & 223.58492 & 5.96500 & 0.1266 & 369 $\pm$ 47 \\
g24 &   29 & 223.98189 & 9.04349 & 0.0492 & 329 $\pm$ 43 \\
g25 &   29 & 223.98324 & 0.94956 & 0.0434 & 208 $\pm$ 27 \\
g26 &   29 & 228.95267 & 3.06713 & 0.0375 & 269 $\pm$ 35 \\
g27 &   29 & 229.94223 & 1.41210 & 0.0522 & 251 $\pm$ 33 \\
g28 &   29 & 226.33770 & 5.05923 & 0.0805 & 223 $\pm$ 29 \\
g29 &   28 & 231.71999 & 4.88078 & 0.0415 & 202 $\pm$ 27 \\
g30 &   27 & 232.88086 & 7.31242 & 0.0342 & 330 $\pm$ 45 \\
g31 &   27 & 232.04636 & 7.98292 & 0.0755 & 215 $\pm$ 29 \\
g32 &   27 & 231.98863 & 5.15568 & 0.0869 & 307 $\pm$ 42 \\
g33 &   26 & 226.56570 & 3.13856 & 0.0427 & 241 $\pm$ 33 \\
g34 &   25 & 229.87594 & 4.29384 & 0.1015 & 347 $\pm$ 49 \\
g35 &   25 & 229.96599 & 1.58380 & 0.0901 & 205 $\pm$ 29 \\
g36 &   25 & 230.98850 & 6.89963 & 0.0768 & 296 $\pm$ 42 \\
g37 &   24 & 232.67681 & 3.44119 & 0.0867 & 233 $\pm$ 34 \\
g38 &   24 & 228.64094 & 1.92073 & 0.0952 & 245 $\pm$ 35 \\
g39 &   24 & 226.00510 & 9.94780 & 0.0378 & 249 $\pm$ 36 \\
g40 &   24 & 228.97296 & 3.63668 & 0.0797 & 303 $\pm$ 44 \\
g41 &   23 & 231.06640 & 7.95250 & 0.0767 & 248 $\pm$ 37 \\
g42 &   23 & 226.68802 & 3.77257 & 0.0352 & 233 $\pm$ 34 \\
g43 &   22 & 231.96218 & 3.50661 & 0.0844 & 302 $\pm$ 46 \\
g44 &   22 & 226.85276 & 8.58546 & 0.0791 & 305 $\pm$ 46 \\
g45 &   22 & 232.81885 & 4.57287 & 0.0389 & 262 $\pm$ 40 \\
g46 &   21 & 231.79646 & 6.26354 & 0.0429 & 200 $\pm$ 31 \\
g47 &   21 & 225.89265 & 8.94561 & 0.0928 & 322 $\pm$ 50 \\
g48 &   21 & 225.78036 & 8.02864 & 0.0893 & 223 $\pm$ 34 \\
g49 &   21 & 228.14151 & 7.92658 & 0.0808 & 201 $\pm$ 31 \\
g50 &   21 & 224.37460 & 8.82592 & 0.0802 & 246 $\pm$ 38 \\
g51 &   20 & 229.57824 & 7.57849 & 0.0768 & 224 $\pm$ 35 \\
g52 &   20 & 226.92806 & 7.11260 & 0.0310 & 262 $\pm$ 41 \\
g53 &   20 & 232.30867 & 7.53492 & 0.0428 & 215 $\pm$ 34 \\
g54 &   20 & 227.90004 & 9.43298 & 0.0344 & 220 $\pm$ 35 \\
g55 &   20 & 230.20846 & 3.56299 & 0.0372 & 223 $\pm$ 35 \\
g56 &   19 & 227.54331 & 1.24429 & 0.0730 & 220 $\pm$ 36 \\
g57 &   19 & 230.21875 & 2.24033 & 0.1121 & 212 $\pm$ 34 \\
g58 &   19 & 224.11556 & 6.46473 & 0.0460 & 217 $\pm$ 35 \\
g59 &   19 & 227.44930 & 8.83889 & 0.0805 & 244 $\pm$ 40 \\
g60 &   18 & 227.41959 & 8.78529 & 0.0779 & 290 $\pm$ 48 \\
g61 &   18 & 230.10103 & 5.17182 & 0.1019 & 215 $\pm$ 36 \\
g62 &   18 & 228.28278 & 4.32300 & 0.0792 & 266 $\pm$ 44 \\
g63 &   18 & 227.50863 & 3.00206 & 0.0924 & 220 $\pm$ 37 \\
g64 &   18 & 232.70451 & 6.73802 & 0.0750 & 221 $\pm$ 37 \\
g65 &   17 & 232.22761 & 1.73884 & 0.0906 & 254 $\pm$ 44 \\
g66 &   16 & 232.36585 & 3.51020 & 0.0377 & 235 $\pm$ 42 \\
g67 &   16 & 231.24717 & 2.80731 & 0.0517 & 227 $\pm$ 40 \\
g68 &   16 & 231.79862 & 1.40464 & 0.1167 & 325 $\pm$ 58 \\
g69 &   15 & 225.92630 & 10.31350 & 0.0948 & 266 $\pm$ 49 \\
g70 &   15 & 225.90523 & 7.74571 & 0.0881 & 292 $\pm$ 53 \\
g71 &   15 & 223.55670 & 5.69663 & 0.0819 & 209 $\pm$ 38 \\
g72 &   14 & 229.71561 & 6.19543 & 0.0989 & 213 $\pm$ 40 \\
g73 &   14 & 228.30915 & 2.80182 & 0.1003 & 370 $\pm$ 70 \\
g74 &   13 & 225.66892 & 4.78330 & 0.1228 & 213 $\pm$ 42 \\
g75 &   13 & 228.24896 & 3.10813 & 0.0295 & 210 $\pm$ 41 \\
g76 &   13 & 225.89818 & 7.83662 & 0.0479 & 219 $\pm$ 43 \\
g77 &   12 & 223.70332 & 5.42359 & 0.0936 & 248 $\pm$ 51 \\
g78 &   12 & 228.99423 & 2.00921 & 0.1038 & 265 $\pm$ 54 \\
g79 &   11 & 224.05415 & 2.99775 & 0.0699 & 207 $\pm$ 44 \\
g80 &   11 & 231.61411 & 0.84675 & 0.1137 & 270 $\pm$ 58 \\
g81 &   11 & 225.10084 & 4.87849 & 0.0725 & 207 $\pm$ 44 \\
g82 &   10 & 230.72365 & 7.33575 & 0.0761 & 310 $\pm$ 70 \\
g83 &   10 & 227.29251 & 0.63658 & 0.0897 & 260 $\pm$ 58 \\
g84 &   10 & 230.49468 & 2.46260 & 0.0825 & 220 $\pm$ 49 \\
g85 &   10 & 228.62163 & 10.23015 & 0.0548 & 225 $\pm$ 50 \\
g86 &   10 & 230.34193 & 5.11939 & 0.0438 & 217 $\pm$ 49 \\
g87 &   10 & 232.01456 & 2.73368 & 0.1172 & 206 $\pm$ 46 \\

\end{longtable}
 
\subsection{X-ray clusters}

We compare our results with the compilation of the ROSAT X-ray cluster catalog -- the second release of the Meta-Catalog of X-ray detected clusters of galaxies \citep[MCXCII,][]{2024AA...688A.187S}. This catalog lists 12 clusters with redshift $0.02<z<0.13$ in this field. Eleven of these clusters coincide with our  structures (Table \ref{tab:x});  MCXC J1521.9+0827 appears to be the only missing X-ray cluster. However, its celestial coordinates (RA$=230.4760$ deg, DEC$=8.4590$ deg) and its redshift ($z=0.036$) are very close to A2063 (RA$-230.75750$ deg, DEC$=8.63944$ deg, $z=0.035$). Therefore, the BT algorithm does not identify  this cluster as a structure distinct from A2063.

The BT algorithm separates A2029 into two structures, g15 and g22, because these two structures, which overlap on the sky, have a large redshift separation $\Delta z \sim$ 0.0062, corresponding to a velocity separation $c\Delta z \sim$ 1860 km~ s$^{-1}$. This velocity separation
is also reported by the X-ray spectral measurement of Suzaku \citep{2016PASJ...68S..19O}. 
The X-ray XRISM Resolve observed the hot gas centered on the BCG of A2029 with a field of view of 3 arcmin, which corresponds to $\sim 270$ kpc at the the distance of A2029. For the hot gas, the XRISM observation measures the redshift $z=0.0779$, which is consistent with the BCG redshift; this result indicates that the gas is nearly at rest in the gravitational potential well of the BCG \citep{2025ApJ...982L...5X}. On the other hand, this redshift is  between the redshifts of the two structures g15, to which the BCG is associated, and g22: so, the dynamical structure of the whole galaxy system appears complex and might suggest that g15 and g22 are currently merging, with g22, with a smaller velocity dispersion and thus probably less massive than g15, falling onto the more massive g15.

Similarly, A2055 is separated into the two BT structures g21 and g72, with a velocity separation 
$c\Delta z \sim$ 1049 km~ s$^{-1}$.
Indeed, both A2055 and A2029 are suggested to be merging systems \citep{2016MNRAS.458..226D}.

The BT algorithm is effective at identifying structures that are known to be X-ray sources but are not listed in catalogs of X-ray clusters:  g11 corresponds to a substructure south of A2029 \citep[labeled as A2029S,][]{2017Sohn,2019Sohn}; g60 corresponds to the X-ray source 1RXS J150935.9+084605 \citep{2010A&A...522A..34G,2012Walker}.
All these associations are listed in Table \ref{tab:x}.

\begin{longtable}{|c|c|c|c|c|c|c|}
\caption{Associations of the BT structures with known clusters}
\label{tab:x} \\
 \hline
 ID & sID & NED & MCXCII & MSPM & GalW \\
\hline
\endfirsthead
\multicolumn{6}{c}{\textbf{continued}} \\
\hline
 ID & sID & NED & MCXCII & MSPM & GalW \\
 \hline
\endhead

\hline
\endfoot

g1 & s5 & A2040 & J1512.7+0725 & 57 & 114 \\
g2 & s1 & A2063 & J1523.0+0836 & 17 & 82 \\
g3 & s6 & MKW03s & J1521.8+0742 & 47 & 262 \\
g4 & s2 & A2052 & J1516.7+0701 & 72 & 278 \\
g5 & s24 & A2048 & J1515.3+0423 & 6036 & 39 \\
g6 & s50 &  &  & 49 & 314 \\
g7 & s46 &  &  & 220 &  \\
g8 &  &  &  &  & 1106 \\
g9 & s10 &  &  & 35 & 1513 \\
g10 & s14 & A2028 &  & 4364 & 481 \\
g11 & s7 & & A2029S  & 4331 & 113 \\
g12 & s15 &  & J1511.5+0145 & 90 & 1534 \\
g13 & s36 &  &  &  & 835 \\
g14 & s3 & A2033 & J1511.3+0619 & 4420 &  \\
g15 & s4 & A2029 & J1510.9+0543 & 3489 & 6 \\
g16 &  &  & J1528.7+0133 & 5922 & 585 \\
g17 & s29 &  &  &  &  \\
g18 & s25 &  &  &  &  \\
g19 & s8 &  &  & 257 &  \\
g20 &  &  &  & 5770 &  \\
g21 & s9 & A2055 & J1518.7+0613 &  & 46 \\
g22 &  & A2029 & J1510.9+0543 & 3489 & 6 \\
g23 & s49 &  &  &  &  \\
g24 &  &  &  &  &  \\
g25 & s71 &  &  & 457 &  \\
g26 & s11 &  &  & 143 &  \\
g27 & s42 &  &  &  &  \\
g28 &  &  &  & 5064 &  \\
g29 &  &  &  & 782 &  \\
g30 & s12 & A2085 &  & 137 &  \\
g31 & s34 &  &  & 5307 & 1267 \\
g32 & s72 &  &  & 6366 &  \\
g33 & s13 &  &  & 102 &  \\
g34 & s32 &  &  &  & 514 \\
g35 &  &  &  &  &  \\
g36 &  &  &  &  & 1234 \\
g37 & s55 & A2082 &  & 4880 & 1095 \\
g38 & s58 &  &  & 6092 &  \\
g39 & s63 &  &  & 1396 &  \\
g40 & s64 &  &  &  &  \\
g41 & s45 &  &  &  &  \\
g42 & s16 &  &  & 634 &  \\
g43 & s69 &  &  & 4428 & 581 \\
g44 &  &  &  &  &  \\
g45 &  &  &  & 33 & 692 \\
g46 & s53 &  &  & 1534 &  \\
g47 &  &  &  &  &  \\
g48 & s57 &  &  & 6060 &  \\
g49 &  &  &  &  & 1104 \\
g50 & s41 &  &  & 4665 &  \\
g51 &  &  &  &  &  \\
g52 & s21 &  &  & 547 &  \\
g53 & s40 &  &  & 46 &  \\
g54 & s20 &  &  & 1009 &  \\
g55 & s19 &  &  & 887 &  \\
g56 &  &  &  &  &  \\
g57 &  &  &  & 8186 &  \\
g58 & s26 &  &  & 701 &  \\
g59 & s22 & J150935&  & 4406 & 838 \\
g60 & s27 & J150935 &  & 4406 & 838 \\
g61 &  &  &  & 7302 & 1051 \\
g62 & s30 &  &  & 4703 &  \\
g63 &  &  &  & 4689 & 671 \\
g64 &  &  &  &  &  \\
g65 & s33 &  &  & 4894 & 1741 \\
g66 & s35 &  &  & 360 &  \\
g67 &  &  &  & 3690 &  \\
g68 &  &  &  &  &  \\
g69 & s37 &  &  & 4648 &  \\
g70 & s38 &  &  & 4567 &  \\
g71 & s59 &  &  & 5048 & 1771 \\
g72 & s44 & A2055 & J1518.7+0613 &  & 46 \\
g73 & s43 &  &  & 4583 &  \\
g74 &  &  &  &  &  \\
g75 &  &  &  & 602 &  \\
g76 & s47 & A2040 &  &  &  \\
g77 & s51 &  &  & 4459 &  \\
g78 & s52 &  &  &  & 579 \\
g79 & s61 &  &  & 3187 &  \\
g80 & s62 &  &  & 7608 &  \\
g81 &  &  &  & 4007 &  \\
g82 & s67 &  &  & 6428 & 1665 \\
g83 &  &  &  &  &  \\
g84 & s66 &  &  &  & 1106 \\
g85 &  &  &  & 2144 & 681 \\
g86 & s70 &  &  &  &  \\
g87 & s68 &  &  & 7506 &  \\

\end{longtable}

\subsection{Optical clusters}

We now compare our list of structures with cluster catalogs based on optical data.
We associate one of the BT structures with an optical cluster when (i) the separation of their centers, projected on the sky, is smaller than 1 Mpc at the redshift of the BT structure, and (ii) their redshift separation is smaller than 0.01.

There are 10 Abell clusters with redshift $z \in [0.02, 0.13]$ in our field \citep{1989ApJS...70....1A}. They all match our BT structures (see Table \ref{tab:x}). For A2020, the NED redshift $z=0.058$ was originally estimated with only one galaxy \citep{1999ApJS..125...35S, 2002AJ....124.1918M}. However, a more recent estimate based on 16 galaxies yields $z=0.048$ \citep{2014Lauer}, making A2020 coincide with our BT structure g76. 

We also consider the cluster catalogs based on the SDSS  spectroscopic data, such as the MSPM catalog from DR7 \citep{2012MNRAS.422...25S}, whose spectra sample is complete to mag$_r= 17.77$, and the GalWeight catalog based on DR13 \citep{2020ApJS..246....2A}, whose data in this field are the same as DR18. 
Within the redshift interval [0.02, 0.13], the GalWeight catalog lists 43 galaxy clusters in this field of view. Twenty-nine of them match the BT structures. The MSPM catalog, based on the multi-scale probability mapping algorithm, identifies 200 cluster/group candidates, 57 of which match our BT structures. These associations are listed in Table \ref{tab:x}.

The BT structures that have no counterparts in existing catalogs can be either newly found galaxy systems or spurious systems, namely systems that originate from a chance alignments of galaxies in redshift space but do not correspond to any actual system in real space.
To identify the BT structures that might be spurious systems, we proceed as follows.
Considering that real clusters usually contain dense cores that can be detected with a large $\Delta \eta$ threshold,
we run the BT algorithm by adopting the threshold $\Delta \eta = 50$.
With this threshold, we identify 72 structures
with velocity dispersions larger than 200 km~s$^{-1}$ and at least 10 members.
We compare these 72 high-density structures with the 87 low-density structures identified above with $\Delta \eta = 20$. Out of the 72 high-density structures, 71 are dense substructures of 61 low-density structures. The second column of Table \ref{tab:x} lists the 71 high-density structures, labeled with the prefix ``s'', on the row of the associated low-density structure.  The missing high-density structure s26 is still a substructure of a low-density structure identified with $\Delta\eta=20$, but this latter structure is not included in the sample of the 87 structures because its velocity dispersion is smaller than 200 km s$^{-1}$.

 \begin{figure}
 \centering
 \includegraphics[width=0.48\textwidth]{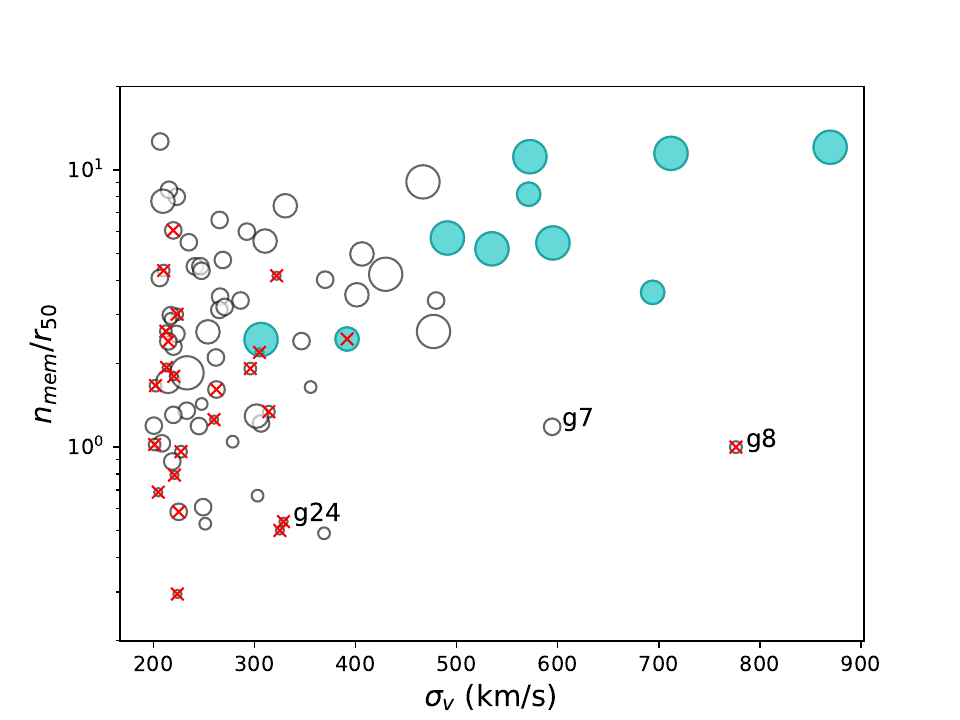}
 \caption{Central density against velocity dispersion of the 87 BT structures.
 The size of each circle is proportional to the number of clusters associated to the structure as listed in Table \ref{tab:x}. BT structures with X-ray emission are indicated by cyan dots. The 26 low-density structures that are missing from the sample of 72 high-density structures 
 are marked with an additional red cross.
 }
 \label{fig:sigman}
\end{figure}

The remaining 26 low-density structures that do not contain any of the 72 high-density structures have a relatively low number of members and a particularly low central density. To illustrate this result, Figure \ref{fig:sigman} shows the quantity $n_{mem} / r_{50}$ as a function of the velocity dispersion $\sigma_v$ of our 87 low-density structures; here, $r_{50}$ is the radius of the circle embracing half of the members of the structure:  $n_{mem} / r_{50}$ thus estimates the density of the central region of the structure.

The 26 structures missing from the sample of the 72 high-density structures, shown as red crosses in Fig. \ref{fig:sigman}, cluster towards the low velocity dispersion  part of the diagram: they might thus be low-mass groups. Alternatively, some or all of them might be spurious systems. Indeed, they represent $\sim 30\%$ of our sample of the 87 low-density BT structures: if these 26 structures were all spurious, this fraction would not be at odds with the $\sim 50\% $ spurious substructures, with velocity dispersion larger than $200$~km~s$^{-1}$, of galaxy clusters that are found in mock catalogs \citep{Yu2018a}. 

It is worth noting that there are two structures, g7 and g8, with large velocity dispersion but still relatively  low central density. Below, we show that we associate these structures to cosmic filaments.

\section{Superclusters}
\label{sec:superclusters}

To identify galaxy clusters and groups, the BT method arranges the galaxies in a hierarchical structure. This feature enables the investigation of larger and less dense systems, like galaxy superclusters and filaments.

\begin{figure}
 \centering
 \includegraphics[width=0.49\textwidth]{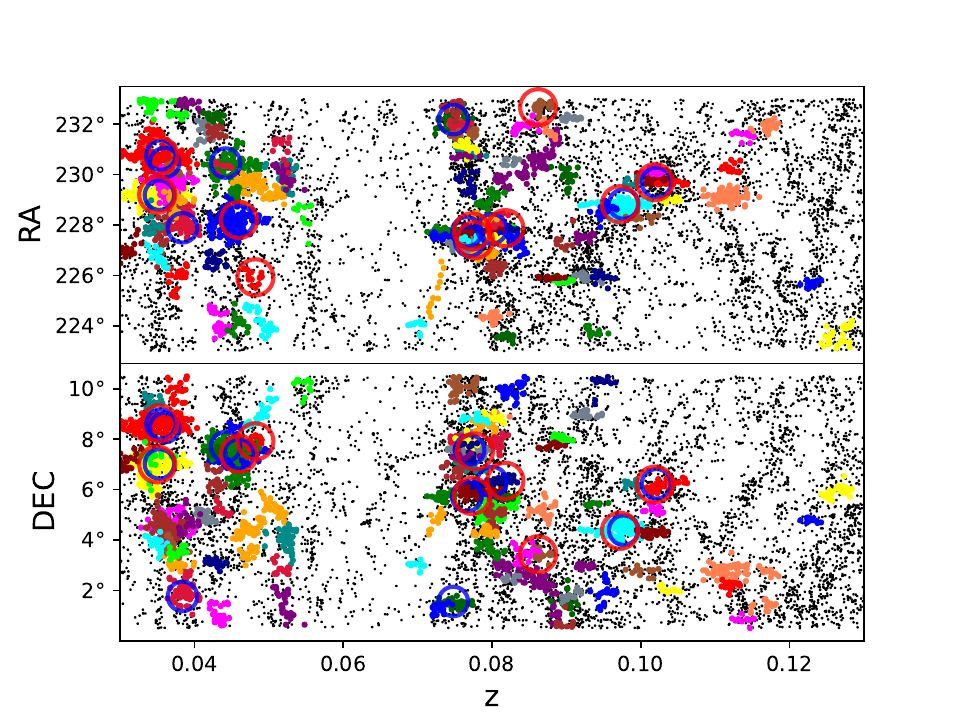}
 \caption{The redshift distribution of the structures  with at least 20 members and velocity dispersion larger than 200 km~s$^{-1}$. The grey points, the blue circles, and the red circles show the galaxies, the MCXC clusters, and the Abell clusters, respectively.}
 \label{fig:zdist}
\end{figure}

The redshift distribution of the BT structures shown in  Fig. \ref{fig:zdist} suggests the presence of a number of superclusters at different redshifts. 
By adopting the low threshold $\Delta \eta = 1$, the BT algorithm can identify the superclusters and its member galaxies.
We identify three superclusters with more than 400 member galaxies.
We list their properties in Table \ref{tab:supercl}.

All these three systems coincide with  known superclusters \citep{2014Chow}.
The supercluster s1, the A2029-A2033 system, is the dominant structure in the field: it includes numerous clusters gathering around A2029. The two superclusters s2, the A2040-MKW03s system, and s3, the A2063-A2052 system, appear rather loose with no obvious central system.
\begin{table}
\caption{Superclusters ($\Delta \eta = 1)$}
\label{tab:supercl}
 \begin{tabular}{|c|c|c|c|l|}
 \hline
 \multirow{2}{*}{ID} &  \multirow{2}{*}{$n_{\rm mem}$} &  \multirow{2}{*}{z} & $\sigma_v$ & \multirow{2}{*}{core clusters} \\
  &       &       & (km~s$^{-1}$)  &   \\
\hline
s1 &  1008 & 0.079 & 785 & A2029-A2033  \\
s2 &  487 & 0.046 & 664 & A2040-MKW03s  \\
s3 &  453 & 0.034 & 566 & A2063-A2052 \\
\hline
 \end{tabular}
\end{table}

\begin{figure}
 \centering
 \includegraphics[width=0.48\textwidth]{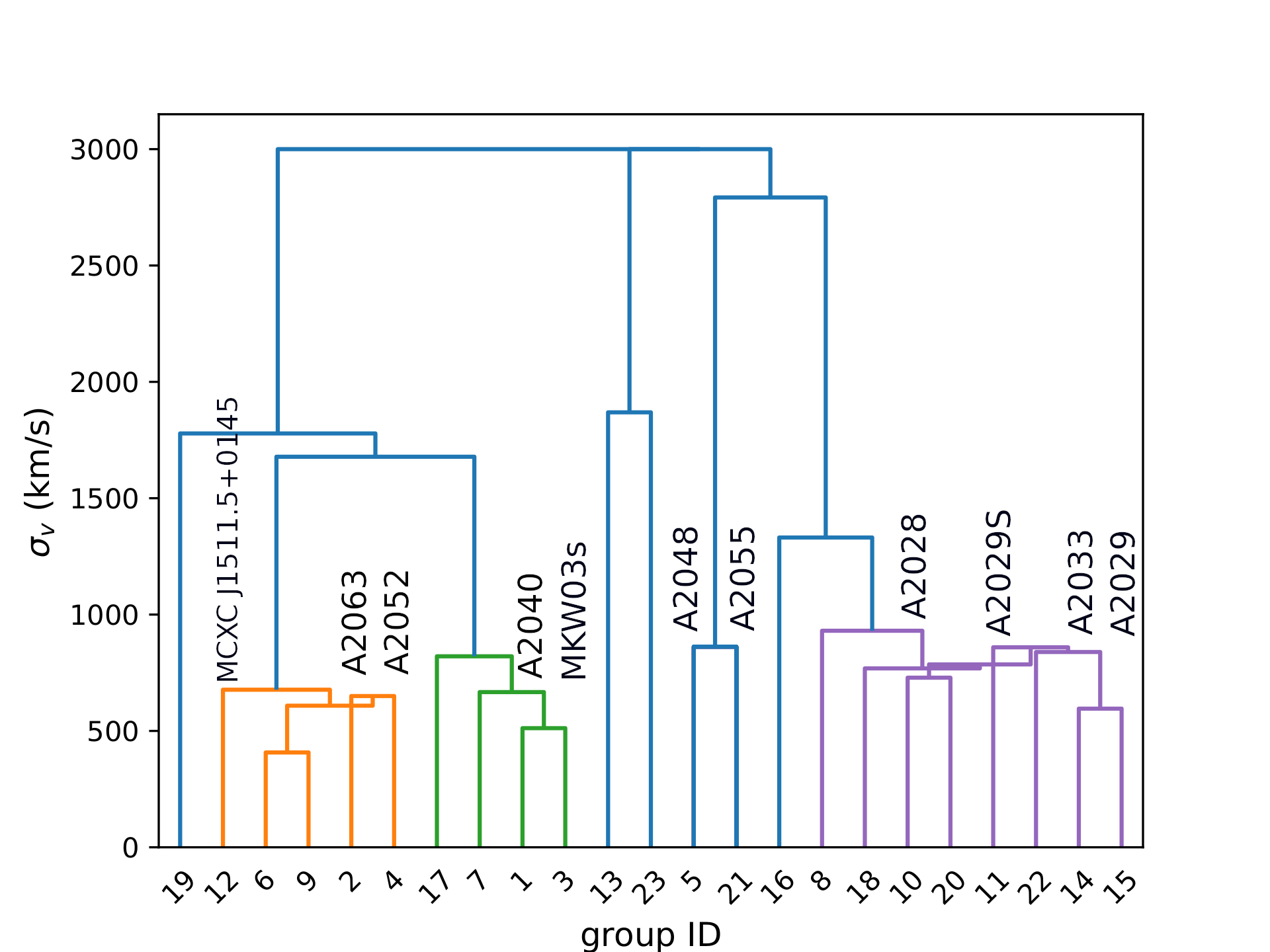}
 \caption{The simplified hierarchical tree of BT structures. Only structures with $\sigma_v > 250 $ km~s$^{-1}$ and $n_{mem} \ge 30$ are shown. The vertical axis shows the velocity dispersion of the nodes. The orange, green, and violet branches show the supercluster s3, s2, and s1, respectively. }
 \label{fig:stree}
\end{figure}

To unveil the hierarchical structure of the superclusters, we build a binary tree by considering each cluster or group as an individual leaf  (Fig. \ref{fig:stree}).
As a cluster or group, we consider all the BT structures identified with $\Delta \eta = 20$ containing at least 30 members, with  velocity dispersion larger than 250 km~s$^{-1}$. The binary tree shown in Fig. \ref{fig:stree} clearly shows how the individual galaxy systems group together to form superclusters.

\subsection{Supercluster s1 (A2029-A2033)}

\begin{figure}
 \centering
 \includegraphics[width=0.48\textwidth, trim={1.5cm 0 1.5cm 0}]{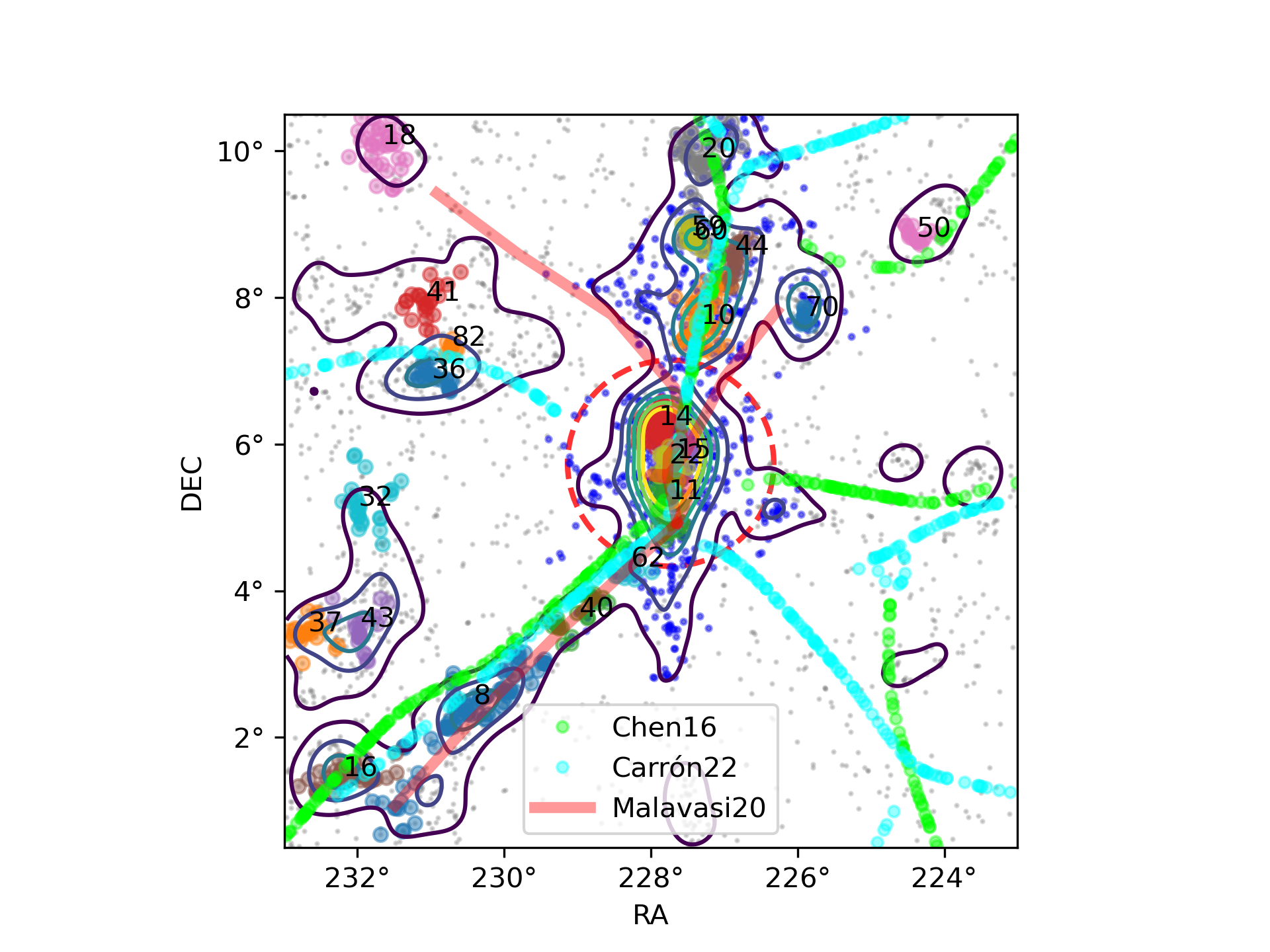}
 \caption{The distribution on the sky of the 1008 member galaxies (blue dots) of the supercluster s1 (A2029-A2033). The contours show the number density of  galaxies with spectroscopic redshift in the range $z=[0.069-0.089]$. The dots with different colors show the BT structures with velocity dispersion $\sigma_v > 230$ km~s$^{-1}$ in this redshift slice. 
 The grey dots are foreground or background galaxies. The red dashed circle has radius 5$R_{200}$ centered on the BCG of A2029, with $R_{200}=1.91$ Mpc the radius of A2029. The colored curves show the filaments suggested in the literature \citep{Chen16,2020Malavasi,2022Duque}. }
 \label{fig:density}
\end{figure}

Figure \ref{fig:density} illustrates the distribution on the sky of the member galaxies in the supercluster s1 (A2029-A2033).
The contour levels, derived from a Gaussian kernel density estimator, show the galaxy number density.
The clusters A2029 (g15, g22), A2033 (g14), and A2029S (g11) appear in  the central region of the supercluster.
Based on the binary tree shown in Fig. \ref{fig:stree}, in addition to identify the spatial distribution of the members of the supercluster, we can assess its hierarchical structure.
The cluster A2029S (the BT structure g11) appears above A2033 in the binary tree; in other words, A2033 is  deeper than A2029S in the gravitational potential well of the supercluster;
it is thus likely that the accretion of A2029S occurred later than A2033. This results is consistent with the conclusion drawn from galaxy spectra by \citet{2019Sohn}.

A2028 (g10) is located in the northern outskirt. It is probably still falling onto the supercluster, as inferred from its cometary-like X-ray emission \citep{2010A&A...522A..34G}.
Three additional structures, g20, g44, and g59, also appear to be falling onto the center along the same filamentary structure.
Similarly, the BT structures g62, g40, and g8, located in the southeast of A2029, show elongated shapes and are aligned along a filamentary structure. This feature is consistent with the results from the catalogs of cosmic filaments \citep{Chen16,2020Malavasi,2022Duque}. We conclude that these BT structures identify two cosmic filaments. In addition, the elongated shape of g8 suggests that it is not a galaxy  group but rather an overdensity of the filament.

\subsection{Supercluster s2 (A2040-MKW03s)}

\begin{figure}[htp]
 \centering
 \includegraphics[width=0.48\textwidth, trim={1.5cm 0 1.5cm 0}]{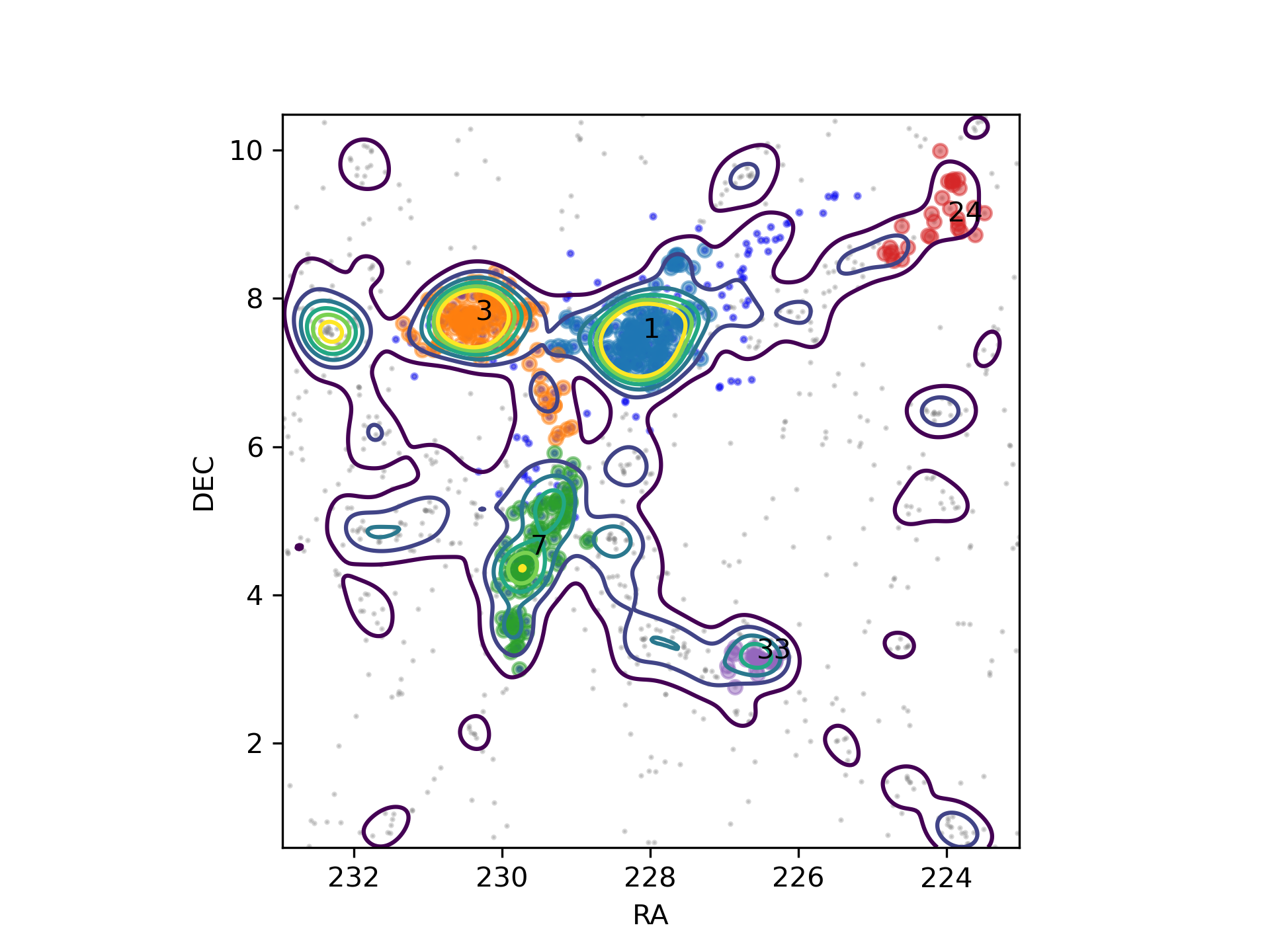}
 \caption{Same as Fig. \ref{fig:density} for the 487   member galaxies of the supercluster s2 (A2040-MKW03s).
}
 \label{fig:s2}
\end{figure}

Figure \ref{fig:s2} shows the  distribution on the sky of the member galaxies of the supercluster s2.
The  contours show the number density of galaxies in the redshift range $z=[0.036, 0.056]$. The BT structures shown in the figure are within the same redshift range.

Supercluster s2 is dominated by the two clusters A2040 (g1) and MKW03s (g3). 
The two BT structures g7 and g24 in the outer region have a shape elongated toward A2040.
Their low central density  (see  Fig. \ref{fig:sigman}) suggests that each of them is an overdensity of a filament.
We are unable to compare our result with other investigations, because catalogs of cosmic filaments within $z=0.05$ are currently unavailable in the literature.

\subsection{Supercluster s3 (A2063-A2052)}

 \begin{figure}
 \centering
 \includegraphics[width=0.48\textwidth, trim={1.5cm 0 1.5cm 0}]{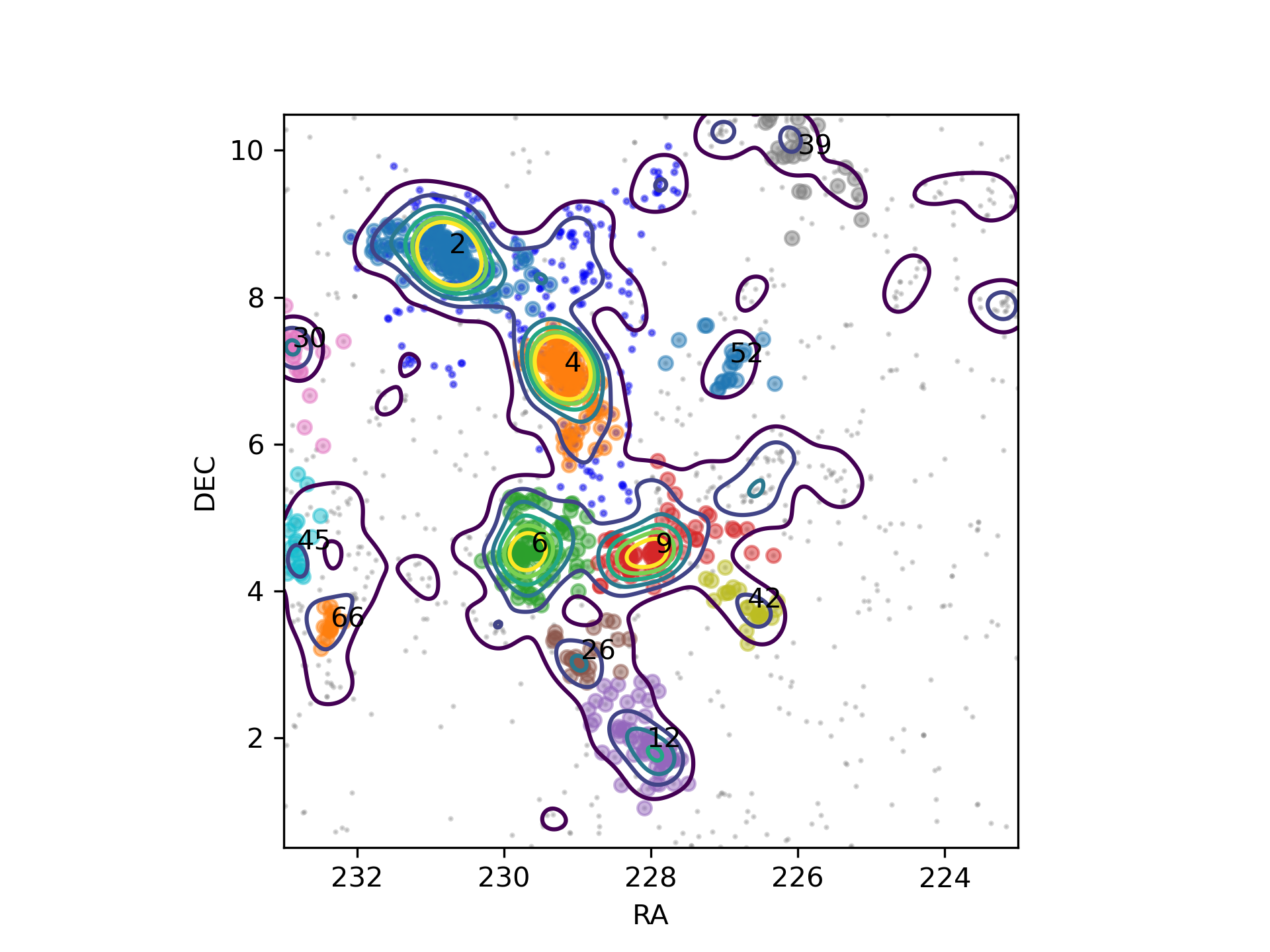}
 \caption{Same as Fig. \ref{fig:density} for the 453 member galaxies of the supercluster s3.
}
 \label{fig:s3}
\end{figure}

The celestial distribution of the member galaxies of the supercluster s3 shows a complex scenario, without a clearly dominant central region (Fig. \ref{fig:s3}).
According to the binary tree shown in Fig. \ref{fig:stree}, g6 and g9 are at the bottom of the gravitational potential well.
The cluster pair A2063 (g2) and A2052 (g4) is falling onto it.
A number of small groups in the outer region will likely fall in later on.

\section{conclusion}
\label{sec:end}
The Blooming Tree (BT) algorithm, based on the  hierarchical clustering method, is  designed for identifying cosmic structures. It provides a unified solution for detecting galaxy groups, clusters, and superclusters based on spectroscopic redshifts. The BT algorithm searches for systems independently of their shape and their dynamical state,
making it particularly suitable for studying galaxy systems which are out of dynamical equilibrium, like merging galaxy clusters, filaments, and superclusters.

In this paper, we apply the BT method to
a large-field-of-view compilation of data from a 100 square-degree sky region around the massive galaxy cluster A2029. The method  identifies clusters and groups of galaxies, including all the known X-ray  galaxy clusters in the field of view.
It identifies additional optical clusters and groups, some of which are unrelaxed, merging with other systems or have filamentary shape.

By lowering the detection threshold $\Delta \eta$, the BT algorithm can identify looser systems, like superclusters. 
The algorithm returns the galaxy membership, and, more importantly,
the hierarchy of the galaxy systems making up the superclusters.
This feature of the algorithm is crucial for assessing the accretion history of the superclusters and the evolution of the cosmic structure.

The accuracy of the BT algorithm is limited by the completeness of the spectroscopic redshift survey. The next generation of spectroscopic redshift surveys, such as the Dark Energy Spectroscopic Instrument \citep[DESI,][]{DESI2024II} and the Nancy Grace Roman Space Telescope \citep{2015Spergel}, will provide entirely new opportunities for the application of this method. Moreover, the results of the BT algorithm can be compared with sensitive X-ray galaxy cluster catalogs from the eROSITA X-ray survey \citep{2012eROSITA}. A systematic comparison of  optical and X-ray galaxy cluster data will increase our understanding of the nature of baryons, which are mostly in their hot phase, and of the dark matter within galaxy clusters and the large-scale structure.

\begin{acknowledgments}

We sincerely thank the reviewer, the statistics editor and the data editor for their valuable suggestions that improved the presentation of our results. We would also like to thank Dr. Xiaoyuan Zhang and Prof. Jiaxin Han for their helpful discussions.
This work has been supported by the National Key Research and Development Program of China (No. 2023YFC2206704)  and the China Manned Space Program with grant No. CMS-CSST-2025-A04.
AD acknowledges partial support from the INFN grant InDark.

This research has made use of the NASA/IPAC Extragalactic Database (NED),
which is operated by the Jet Propulsion Laboratory, California Institute of Technology,
under contract with the National Aeronautics and Space Administration, and of NASA’s Astrophysics Data System Bibliographic Services.

\vspace{5mm}
\facilities{SDSS, Guoshoujing Telescope (LAMOST)}

\software{astropy \citep{2013A&A...558A..33A},  scipy \citep{2020SciPy-NMeth}, scikit-learn \citep{scikit-learn} }

\end{acknowledgments}

\appendix

\section{Velocity dispersion error estimate}
\label{app:bootstrap}

To estimate the 1$\sigma$ confidence level of the velocity dispersion, we perform a test with the bootstrap method \citep{Ivezic2014}.
We randomly sample $n$ times a normal distribution with mean $\mu_t$=0 and standard deviation $\sigma_t$.
We then estimate the standard deviation $\sigma_v$ of these $n$ values. We repeat this procedure 1000 times and obtain a distribution of the deviations $\sigma_v$ and their standard deviation $\Delta\sigma_v$.

\begin{figure}
 \centering
 \includegraphics[width=0.45\textwidth]{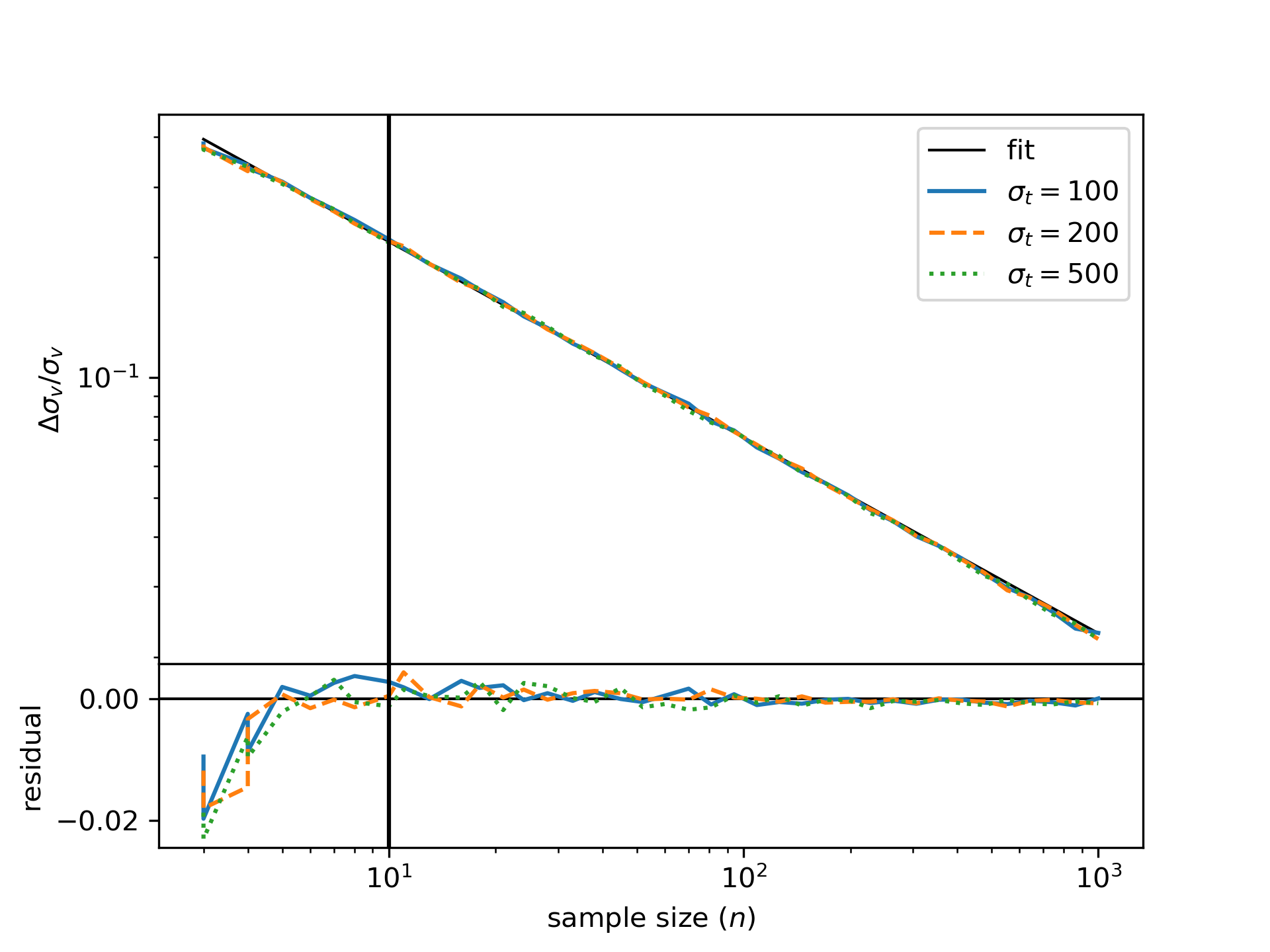}
 \caption{The relation between the sample size $n$ and the standard deviation with standard error estimation $\Delta \sigma_v$. The three lines correspond to different values of $\sigma_t$. The black solid line is the fitted relation.
 }
 \label{fig:vdis_e}
\end{figure}

We explore the range $n=[3,1000]$.
Figure \ref{fig:vdis_e} shows that the relation between $\Delta\sigma_v/\sigma_v$ and $n$ is described by the linear relation
\begin{equation}
{\rm log_{10}}\left( \frac{\Delta \sigma_v}{ \sigma_v}\right) = -0.49 ~{\rm log_{10}}n - 0.17\; .
\label{eq:vdis_e}
\end{equation}
We test that this linear relation holds for $\sigma_t=100, 200, 500$.
When the sample size is above $n=10$, $\Delta\sigma_v/\sigma_v$ is smaller than 22\%.

By identifying the Gaussian distribution as the distribution of the velocities of the member galaxies of the BT structures, and $\sigma_t$ as their true velocity dispersion, we can adopt $\Delta\sigma_v$ as the relative uncertainty on  the velocity dispersion $\sigma_v$ estimated  with $n$ members. We can thus use eq. (\ref{eq:vdis_e}) for our BT structures. 
 
\bibliography{a2029}

\bibliographystyle{aasjournal}

\end{document}